\documentclass[manuscript,review=false]{acmart}

\usepackage{graphicx}
\definecolor{coBlue}{HTML}{286BA5}
\definecolor{coPurple}{HTML}{7952AA}
\definecolor{coInk}{HTML}{222222}
\definecolor{coSlate}{HTML}{223548}
\makeatletter
\@namedef{co@fig@A}{fig:system}\@namedef{co@fig@B}{fig:details}%
\@namedef{co@fig@C}{fig:receipt}\@namedef{co@fig@D}{fig:usage}%
\@namedef{co@col@A1}{coBlue}\@namedef{co@col@A2}{coBlue}\@namedef{co@col@A3}{coBlue}%
\@namedef{co@col@A4}{coPurple}\@namedef{co@col@B}{coSlate}%
\def\co@parse#1#2\@nil{%
  \@ifundefined{co@col@#1#2}%
    {\@ifundefined{co@col@#1}{\def\co@c{coInk}}{\edef\co@c{\csname co@col@#1\endcsname}}}%
    {\edef\co@c{\csname co@col@#1#2\endcsname}}%
  \edef\co@t{\csname co@fig@#1\endcsname}\def\co@l{#1#2}}
\newcommand{\co@box}[2]{{\setlength{\fboxsep}{1.1pt}\smash{\colorbox{#1}{\textcolor{white}{\sffamily\bfseries\footnotesize #2}}}}}
\def\co@link#1{\hyperref[#1]{\co@box{\co@c}{\co@l}}}
\DeclareRobustCommand{\co}[1]{\co@parse#1\@nil\expandafter\co@link\expandafter{\co@t}}
\DeclareRobustCommand{\cob}[1]{\co@parse#1\@nil\co@box{\co@c}{\co@l}}
\makeatother
\acmConference[IUI Companion '27]{32nd International Conference on Intelligent User Interfaces}{February 8--11, 2027}{Helsinki, Finland}
\acmBooktitle{32nd International Conference on Intelligent User Interfaces (IUI Companion '27), February 8--11, 2027, Helsinki, Finland}
\acmYear{2027}
\acmDOI{}
\acmISBN{}

\setcopyright{none}
\renewcommand\footnotetextcopyrightpermission[1]{}
\begin{document}

\title[Project Progress Desk]{Project Progress Desk: Know Where All Your Claude Code and Codex Projects Stand at a Glance}

\newcommand{\au}[2]{\mbox{#1\textsuperscript{#2}}}
\makeatletter
\def\@typeset@author@line{%
  \andify\@currentauthors\par
  {\raggedright\noindent\@currentauthors\par}%
  \def\@currentauthors{}%
  \ifx\@currentaffiliations\@empty\else
    \andify\@currentaffiliations
    {\vspace{0.6ex}\raggedright\noindent\@currentaffiliations\par}%
  \fi
  \def\@currentaffiliations{}}
\makeatother
\author{\au{Wanghao Ye}{1}, \au{Yiting Wang}{1}, \au{Shwai He}{1}, \au{Guoheng Sun}{1}, \au{Ziyao Wang}{1},
  \au{Yexiao He}{1}, \au{Zheyu Shen}{1}, \au{Zhaoyi Liu}{1}, \au{Bowei Tian}{1}, \au{Daeun Jung}{1},
  \au{Aditya Sarkar}{1}, \au{Parastoo Pilevar}{1}, \au{Xiaochuan Gong}{1}, \au{Yancheng Zhu}{1}, \au{Yingying Fan}{1},
  \au{Qiong Wang}{1}, \au{Yuning Zhang}{1}, \au{Meng Feng}{1}, \au{Siyuan Peng}{1},
  \au{Zhenle Duan}{1}, \au{Sihan Chen}{2}, \au{Ziyi Wang}{3}, \au{Yilong Dai}{4},~\NoCaseChange{and}~\au{Ang Li}{1,*}}
\email{wy891@umd.edu}
\affiliation{\institution{\unskip
  \mbox{\textsuperscript{1}University of Maryland, College Park, USA}\quad
  \mbox{\textsuperscript{2}Carnegie Mellon University, USA}\quad
  \mbox{\textsuperscript{3}Texas A\&M University, USA}\quad
  \mbox{\textsuperscript{4}The University of Alabama}}\city{College Park}\state{Maryland}\country{USA}}
\authorsaddresses{\textsuperscript{*}Corresponding author: Ang Li, angliece@umd.edu, University of Maryland, College Park, USA}
\renewcommand{\shortauthors}{Ye et al.}

\begin{abstract}
Agentic coding tools such as Codex and Claude Code let developers work on many projects in parallel, but keeping track of them still means switching between chats, terminal sessions and providers. We present Project Progress Desk, a local desktop interface that brings overall progress, blockers and pending decisions into one view. Users can see which projects need attention and what remains, and compare earlier approaches before choosing how to continue. Agents report each result in a short receipt; the Desk checks it against the project's checklist and moves the bar, showing earlier estimates in a lighter shade. We describe daily use across 36 projects, including a choice between two implementations based on their recorded outcomes. At the demo, attendees will explore 30 simulated projects, trace completed work to overall progress, and watch a coding agent's new result update the overview.
\end{abstract}



\begin{teaserfigure}
  \centering
  \includegraphics[width=\linewidth]{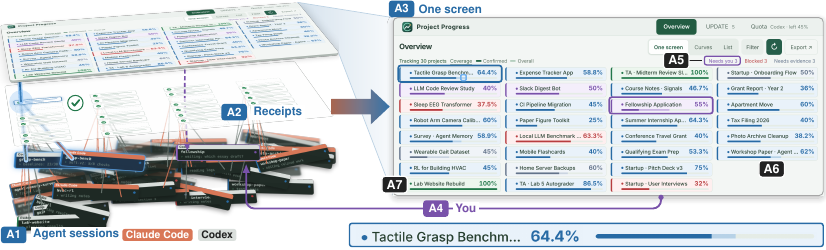}
  \caption{Project Progress Desk. \cob{A1} Coding agents work across projects and sessions. \cob{A2} Agents report results in receipts; the Desk checks each against the project's checklist (\checkmark) and updates progress. Dashed slots have no receipt yet. \cob{A3} The overview, with the blue row enlarged below: a Codex receipt reporting 23 of 30 baseline runs raises this project's progress from 58.4\% to 64.4\%. \cob{A4} The user responds to pending decisions (purple) in the agent's chat. \cob{A5} Counts of projects that need attention. \cob{A6} Dark bars show progress backed by receipts; the light part is an earlier, dated estimate. \cob{A7} Status colors. Simulated demo data.}
  \Description{On the left, an exploded three-dimensional view. At the bottom, labeled A1 Agent sessions, about forty dark agent windows lie in a messy, overlapping pile at different heights and angles. Each window has a colored title bar naming its agent, orange for Claude Code and light gray for Codex, as a small legend shows, and several projects have one window of each. Above the pile are two neat, tilted layers that share the same grid of 30 project positions in four columns. The middle layer, a translucent white sheet labeled A2 Receipts, has one slot per project: a white receipt card with a small green check where the project has a receipt, and an empty dashed slot where it has none. The top layer is a screenshot of the overview. Thin lines, orange from Claude Code windows and gray from Codex windows, fade as they run from every window up to its project's slot, so the pile resolves into the grid; the back of the pile is slightly blurred. Two windows for the project grasp-bench, a Codex window reading baselines 23 of 30 runs and a Claude Code window reading soft-v2 8 of 8 checks, are highlighted in blue; both blue lines meet at the project's receipt card, which is outlined in blue with a large green check mark beside it, and a blue line continues to the same row in the overview. A dashed purple line runs from the purple row Fellowship Application down to a Codex window, fellowship, waiting: which essay draft? A light band connects the top layer to its flat, large copy on the right, labeled A3 One screen: the Project Progress overview with 30 rows, where the row Tactile Grasp Benchmark at 64.4 percent is outlined in blue with the newly added part of its bar marked, and the counter Needs you 3 and the row Fellowship Application are outlined in purple. A magnified copy of the blue Tactile Grasp Benchmark row appears below the overview on a single line, level with the A1 label: its name and 64.4 percent value, followed by its two-tone bar. Callout A6 sits under the bar of the row Workshop Paper, whose bar is dark for most of its length and light at the end. A purple arrow labeled A4 You runs from the bottom of the screen back to the fellowship window in the pile.}
  \label{fig:system}
\end{teaserfigure}

\maketitle

\section{Introduction}

The time horizon of coding agents, the length of task they can finish on their own, has doubled about every seven months~\cite{kwa2025measuring}; agents now carry out whole chip-design flows~\cite{wang2025mcp4eda}, so one developer can run many projects in parallel. Project records are spread across chats, terminal sessions and branches, often on several providers and accounts. Switching accounts when a quota runs out adds another place to check. After a few days away, it is hard to recall a project's progress, the approaches tried, or which agents need an answer.

People want to leave agents running while they do other work and review the results afterwards~\cite{grunde2026overseeing,faas2026design}. This asynchronous oversight~\cite{kitano2026managing} keeps the human on the loop, watching high-level indicators and stepping in when needed~\cite{wulf2025architecting}. Knowledge workers face a similar task-resumption problem after interruptions~\cite{mark2005notask}, which earlier desktop systems addressed by grouping windows and files by task~\cite{dragunov2005tasktracer,oliver2006swish,shen2009detecting}; a recent one maps files by meaning~\cite{majmudar2026filescape}.

\begin{figure}[t]
  \centering
  \includegraphics[width=\linewidth]{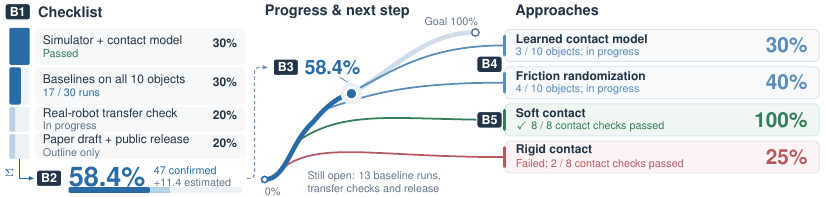}
  \caption{Project details before the update in Figure~\ref{fig:system}~\co{A3}, redrawn from the interface. \cob{B1} Item height shows weight, and fill shows how much is done. \cob{B2} The total adds 47 points backed by receipts to 11.4 points from an October 30 agent estimate for transfer and release work. \cob{B3} The main marker shows the total, 58.4\%. \cob{B4} Branches connect approaches to outcomes, including \cob{B5} eight passed soft-contact checks. Branch geometry is schematic. Simulated demo data.}
  \Description{A compact three-part vector diagram. The left checklist has segment heights proportional to weights of 30, 30, 20 and 20 percent (B1), with dark completion fills and pale estimated fills. The simulator item has passed and baselines have 17 of 30 runs. Transfer is marked in progress and the release item has only an outline; both retain fractions from an earlier agent estimate, not new measured completion. Their weighted sum gives 58.4 percent overall (B2), comprising 47 confirmed and 11.4 estimated points. A dashed connector leads to the same value on the blue main worldline (B3). Its pale continuation leads to the 100 percent goal. Four curved branches connect directly to their approach records (B4): learned contact, 3 of 10 objects, 30 percent; friction randomization, 4 of 10 objects, 40 percent; soft contact, all 8 checks passed, 100 percent (B5); rigid contact, failed with 2 of 8 checks passed, 25 percent. Beneath the curves, remaining work is identified as 13 baseline runs, transfer checks and release. Branch geometry is schematic.}
  \label{fig:details}
\end{figure}

CliDeck groups agent sessions by project~\cite{clideck2026}, and Agent Cockpit gathers their approvals, diffs and history in one interface~\cite{agentcockpit2026}. Human-in-the-loop interfaces help users steer agents~\cite{zhao2026agentgui,mozannar2025magentic,kitano2026managing,zhuang2026agentclick,cheng2026mapping}, debug their runs~\cite{epperson2025agdebugger}, compare trajectories~\cite{yang2026evalagent} and judge outputs against criteria~\cite{desmond2024evalullm,chiang2026multeval}; instructor dashboards show which programming students are struggling~\cite{george2024assistant} and how they ask GPT for help~\cite{george2024notebookgpt}. Jira and GitHub Projects track issues; the Desk ties each session's outcomes to a shared project checklist, so users can see how finished work moves the whole project forward.

We built Project Progress Desk to see, across about three dozen projects, how far along each one is and which ones need our attention. Bar length and color are preattentive cues that support a quick scan~\cite{healey2012attention}, and keeping outcomes and next steps on screen gives users an external memory~\cite{scaife1996external} (Figure~\ref{fig:system}~\co{A3}). We developed the Desk through autobiographical design~\cite{neustaedter2012autobiographical} and will open-source it upon acceptance. This demo contributes:
\begin{itemize}
  \item an overview that gathers agent work from different tools and sessions under each project's goal, with overall progress next to the decisions waiting on the user (Figure~\ref{fig:system}~\co{A3}--\co{A4});
  \item receipts that move each project's bar against its checklist while keeping estimates separate (Figure~\ref{fig:receipt}~\co{C4}--\co{C5});
  \item an approach history that keeps what worked and what failed next to the work still open, so users can decide how to continue (Figure~\ref{fig:details}~\co{B1},~\co{B4}).
\end{itemize}

\section{Project Progress Desk}

\subsection{Overview}

As Shneiderman recommends~\cite{shneiderman1996eyes}, the Desk shows an overview first and details on demand. In a compact dashboard layout~\cite{sarikaya2019dashboards}, the overview fits three dozen projects on a laptop screen, one row each with a name, percentage and bar (Figure~\ref{fig:system}~\co{A3}). Dark bars show progress backed by agent receipts; the light part is an agent's earlier estimate, kept with its date so that its uncertainty stays visible~\co{A6}~\cite{prabhudesai2023understanding,thomas2024assessing,nourani2026heart,gajcin2026cure}. Status colors mark blockers, pending decisions and missing evidence~\co{A7}. Clicking a counter~\co{A5} narrows the list to those projects, and hovering over a row shows its blocker, its next step and the date of its latest evidence. The header shows how much quota each account has left.

\subsection{Progress from Agent Receipts}

\begin{figure}[t]
  \centering
  \includegraphics[width=\linewidth]{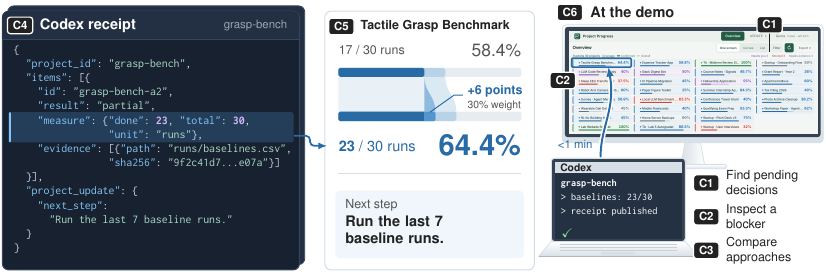}
  \caption{\cob{C4} A Codex receipt links reported work to a project, checklist item, supporting evidence and next step. \cob{C5} Six additional baseline runs move overall progress from 58.4\% to 64.4\%. \cob{C6} At the demo: \cob{C1} find pending decisions, \cob{C2} inspect a blocked project, and \cob{C3} compare approaches. Simulated demo data.}
  \Description{Three groups. Left, labeled C4: a dark navy Codex receipt example shows the project and checklist item identifiers, partial result, a highlighted measure of 23 completed runs out of 30, a supporting file path and abbreviated hash, and the next step Run the last 7 baseline runs. An arrow leads to the updated progress row. Middle, labeled C5: the Tactile Grasp Benchmark panel has two aligned progress bars sharing a zero-to-100-percent scale. The upper row shows 17 of 30 baseline runs and 58.4 percent overall progress; the lower row emphasizes 23 of 30 runs and 64.4 percent. Smooth bands connect the receipt-supported and estimated portions between the rows. The newly added six percentage points are highlighted and annotated with the item's 30-percent weight. Receipt-supported progress increases from 47 to 53 points, while the earlier estimate contributes 11.4 points to both bars. Below, the next step reads Run the last 7 baseline runs. Right, labeled C6: a silver monitor displays the 30-project overview, with the updated project outlined in blue. A laptop shows the Codex terminal reporting the receipt. A blue arrow labeled under one minute connects the laptop to the updated row. Labels C1, C2 and C3 identify finding pending decisions, inspecting a blocker, and comparing approaches.}
  \label{fig:receipt}
\end{figure}

The Desk links each agent session to its project~(Figure~\ref{fig:system}~\co{A1}). A coding agent drafts each project's goal and a weighted checklist, its definition of done; users can inspect both. When an agent finishes a task, it writes a receipt, a provenance record of the result, supporting files, any blocker and the next step (Figure~\ref{fig:receipt}~\co{C4}). The Desk accepts a receipt only if its items are on the current checklist and its files sit inside the project and match their hashes; it then updates the overview within a minute~(Figure~\ref{fig:system}~\co{A2}). If a linked file later changes, the Desk keeps the last valid value and flags the project.

Each item carries a share of the project, and the bar adds up how much of each item is done (Figure~\ref{fig:details}~\co{B1}--\co{B2}). A passed item counts in full, a partial result like 17 of 30 runs counts in proportion, and work no receipt has measured yet keeps its earlier estimate. The bar reaches 100\% only when every item has passed and the project is accepted.

In Figure~\ref{fig:receipt}~\co{C4}--\co{C5}, a Codex receipt raises the baseline runs from 17 to 23 of 30. Baselines make up 30\% of the project, so the six new runs add six points. Users can open the checklist and supporting files to trace each point to its result.

To add a project, the Desk writes a short note into its AGENTS.md, which both Codex and Claude Code read; no plugin is needed. The note asks agents to report what they actually counted, such as 23 of 30 runs, from logs and tests their work already produced, and to keep credentials and raw data out. A separate agent writes the checklist, so the agents doing the work only report against it. Receipts stay on the user's machine.

\subsection{Approaches and Evidence}

Seeing an agent's work helps users understand its behavior~\cite{brachman2025building} and check its output~\cite{zhou2026improving,wu2026stepmind,geyer2026qux360}, but reading through it takes effort~\cite{pareek2026sensemaking,grunde2026overseeing,anik2026designing,nogueira2026verification}. To help users choose the next step, the details view shows earlier approaches (Figure~\ref{fig:details}~\co{B4}), the evidence behind the bar, and links back to the Codex conversations where work can continue.

The main line~\co{B3} tracks overall progress, and each branch~\co{B4}, as in version control, is an approach that an agent tried in an earlier session, ending in its outcome~\cite{ding2025structuring}. In the example, soft contact has passed all eight checks~\co{B5}, but the checklist still lists unfinished baselines, transfer and release work.

\subsection{Daily Use}
\begin{figure}[t]
  \centering
  \includegraphics[width=\linewidth]{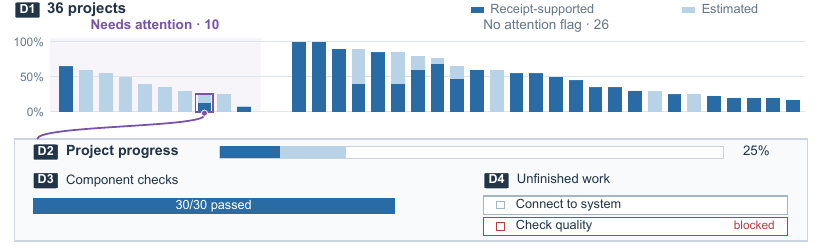}
  \caption{\cob{D1} A 36-project workspace on November 3, 2026, with the highlighted project's \cob{D2} overall progress, \cob{D3} component checks and \cob{D4} unfinished work. Component records are from October 29.}
  \Description{At the top, labeled D1, 36 real projects appear as stacked bars on a zero-to-100-percent axis. Dark blue denotes progress supported by receipts and pale blue denotes estimates. The first group contains ten projects with at least one attention flag; the second contains 26 without an attention flag. Among the ten needing attention, two had entirely receipt-supported progress, one mixed receipts and estimates, and seven had entirely estimated progress. Projects are ordered by overall progress within each group. A purple outline highlights one bar at 25 percent, and a curved connector leads to a detail panel below. Across the top of this panel, Project progress, labeled D2, is filled to 25 percent: 12 points supported by receipts and 13 estimated points. Below it, Component checks, labeled D3, and Unfinished work, labeled D4, form two columns. The component-check bar is full and labeled 30/30 passed, reporting simulation checks from the retained October 29 record. At the right, Connect to system and Check quality have unfilled task boxes; Check quality is explicitly marked blocked in red. The two task rows show selected unfinished requirements. The figure is an anonymized reconstruction combining the November 3 project snapshot with component records from October 29.}
  \label{fig:usage}
\end{figure}

We began using the Desk in September 2026, and about 30 lab members now use it. Figure~\ref{fig:usage}~\co{D1} shows one of our workspaces: of its 36 projects, ten were blocked or waiting on a decision.

We use the Desk to review overnight work and to pick projects back up after time away. Instead of digging through old chats, we open a project and see what is left, what is waiting on us, and what we already tried.

In the highlighted hardware project, the approach history showed that one of two implementations met the timing requirement and the other did not, so we kept the one that met it. The project stood at 25\%~\co{D2}: all 30 component checks passed~\co{D3}, but integration and configuration were unfinished and quality testing was blocked by a storage issue~\co{D4}.

\section{Demonstration}
\label{sec:demo}

In a 3--5-minute session, attendees find projects awaiting their decision~\co{C1}, inspect a project blocked by an expired token~\co{C2}, compare earlier approaches~\co{C3}, and give instructions in linked agent conversations~(Figure~\ref{fig:system}~\co{A4}). A coding agent then reports new work, and the Desk updates the project's bar, status and next step within a minute. The demo runs on a Windows laptop with an external monitor and 30 simulated projects (Figure~\ref{fig:receipt}~\co{C6}).

\section{Discussion and Next Steps}

Progress reflects each project's weighted checklist, with estimated portions distinguished from receipt-supported outcomes. Users can inspect the supporting files to assess reported results. Our experience so far comes from daily use in our lab. Next, we plan a wider deployment and a controlled comparison with a conventional project table, looking at how users find blocked projects and choose what to work on next.

\begin{acks}
We used Codex and Claude to assist with system development and manuscript preparation, and Gemini for manuscript review. The video narration was generated using Chatterbox. We reviewed all content and take responsibility for it.
\end{acks}

\clearpage
\bibliographystyle{ACM-Reference-Format}
\bibliography{refs}

\end{document}